\documentclass[sigconf]{acmart}

\AtBeginDocument{%
  }

\setcopyright{acmlicensed}
\copyrightyear{2026}
\acmYear{2026}
\setcopyright{cc}
\setcctype{by}
\acmConference[SIGIR '26]{Proceedings of the 49th International ACM SIGIR Conference on Research and Development in Information Retrieval}{July 20--24, 2026}{Melbourne, VIC, Australia}
\acmBooktitle{Proceedings of the 49th International ACM SIGIR Conference on Research and Development in Information Retrieval (SIGIR '26), July 20--24, 2026, Melbourne, VIC, Australia}
\acmDOI{10.1145/3805712.3809953}
\acmISBN{979-8-4007-2599-9/2026/07}

\usepackage{amsmath}
\usepackage{amsfonts}
\usepackage{hyperref}
\usepackage{url}
\usepackage{chatonotes}
\usepackage{adjustbox}
\usepackage{booktabs}
\usepackage{tikz}
\usepackage{float}

\newcommand{\para}[1]{\paragraph{\textnormal{\textbf{#1}}}}

\usepackage{enumitem} 
\newcommand{\uls}{\begin{itemize}[leftmargin=*]}
\newcommand{\ule}{\end{itemize}}
\newcommand{\ols}{\begin{enumerate}[leftmargin=*]}
\newcommand{\ole}{\end{enumerate}}
\newcommand{\li}{\item}

\begin{document}

\title{Optimal Re-Ranking Depth}

\author{Siqing Huo}
\affiliation{%
  \institution{University of Waterloo}
  \city{Waterloo}
  \state{Ontario}
  \country{Canada}
}
\email{siqing.huo.canada@gmail.com}

\author{Andrew Parry}
\affiliation{%
  \institution{University of Glasgow}
  \city{Glasgow}
  \country{United Kingdom}
}
\email{a.parry.1@research.gla.ac.uk}

\author{Debasis Ganguly}
\affiliation{%
  \institution{University of Glasgow}
  \city{Glasgow}
  \country{United Kingdom}
}
\email{debforit@gmail.com}

\author{Charles L. A. Clarke}
\affiliation{%
  \institution{University of Waterloo}
  \city{Waterloo}
  \state{Ontario}
  \country{Canada}
}
\email{claclark@gmail.com}

\renewcommand{\shortauthors}{Siqing Huo, Andrew Parry, Debasis Ganguly, \& Charles L. A. Clarke}


\begin{abstract}
Second-stage neural rankers are commonly applied with a fixed re-ranking depth, assuming that retrieval effectiveness saturates as depth increases. Prior research has questioned the assumption that increasing the re-ranking depth yields linear performance gains, further suggesting that optimal re-ranking depth varies considerably from query to query. In the past, studying such phenomena was methodologically difficult given the scale of manual annotation required. With the advent of LLM-based relevance judgments we can now more easily undertake such studies, in this case to pinpoint the optimal re-ranking depth on a per-query basis.
Using dense LLM-based relevance judgments over a typical re-ranking pipeline, we show that many queries exhibit a well-defined optimal re-ranking depth, beyond which effectiveness stagnates or degrades.
We formulate re-ranking depth as a query-specific property and study whether it can be predicted \emph{a priori} from first-stage retrieval characteristics. Through a large-scale analysis, we find that most standard query performance prediction (QPP) methods are ineffective for this task. In contrast, a predictor derived from LLM-assessed first-stage ranking quality, which we term \emph{IR-DCG@10}, can reduce the average re-ranking depth by up to a factor of 3 while preserving overall effectiveness, depending on the specific first- and second-stage rankers used.
Under oracle selection of optimal depths, we further show that retrieval effectiveness can improve by more than 7\% while reducing the average re-ranking depth by a factor of 5 on the MSMARCO DEV collection. Given our promising preliminary findings, we would encourage the use of automatic judgments to facilitate research otherwise infeasible under manual annotation. Towards this point, we release relevance judgments, our codebase, and experimental artefacts to support reproducibility and further research.
\end{abstract}

\maketitle

\section{Introduction}

\begin{figure}[t]
\centering
\includegraphics[width=0.45\textwidth]{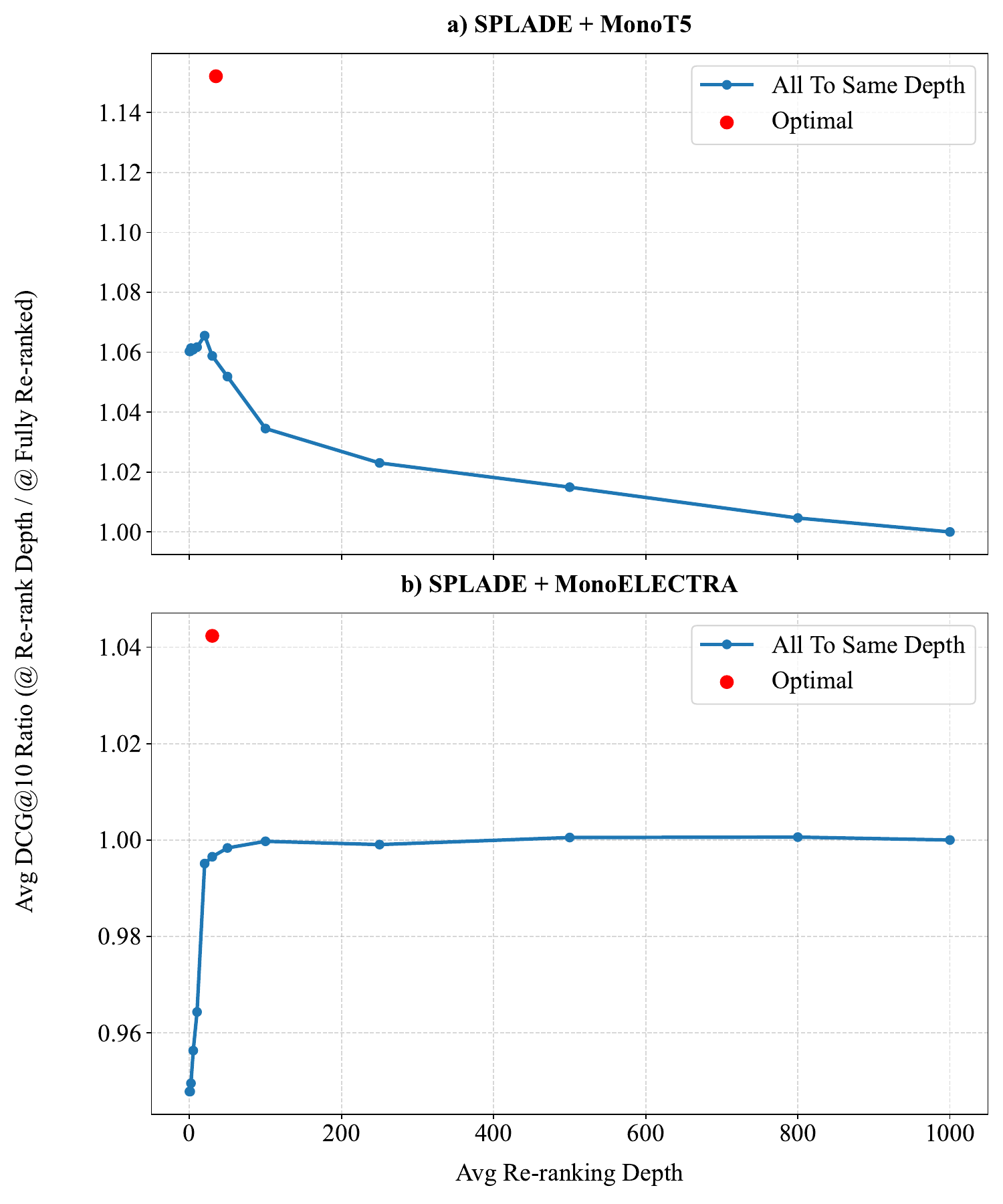}
\caption{The blue lines show the average ratio of each query’s DCG@10 at re-ranking depth $x$ to its DCG@10 when fully re-ranked (depth 1000). The red dots indicate the average optimal re-ranking depth across all queries versus the average ratio of each query’s maximum attainable DCG@10 to its fully re-ranked DCG@10 (depth 1000).
Predicting the optimal ranking depth on a {\em per query} basis would improve both effectiveness and efficiency, reducing average re-ranking depth.}
\label{fig:fixed_depth}
\vspace*{-\baselineskip}
\end{figure}



The objective of an IR system should not only be retrieving documents potentially relevant to an information need but also correctly estimating when, given a user-formulated query, adequate retrieval cannot be achieved with standard ranking procedures, i.e., \emph{knowing when it does not know}. A system that can detect when additional computational effort (such as deeper retrieval or more expensive rankers) is unlikely to improve effectiveness may then engage alternative actions, including asking users to clarify their information need~\cite{aliannejadi2019asking,zamani2020clarifying,mo2023convgqr} or suggesting alternative query formulations~\cite{jiang2018rin,hred_attention,HassanAwadallah,Muntean:2013,querysuggestion,DBLP:conf/cikm/PalumboDWLFFFSB23}.

This capability has particular relevance in modern agentic retrieval workflows, where autonomous agents coordinate multiple reasoning, retrieval, and synthesis steps to satisfy complex information objectives. In Agentic Information Retrieval (Agentic IR), an agent connects observation, reasoning, and action across multiple steps to achieve the desired information state rather than simply returning a static ranked list~\cite{zhang2024agenticIR}. Similarly, multi-agent RAG frameworks orchestrate planners, reformulators, extractors, and synthesizers to address distinct subproblems within a complex task, invoking specialized agents on demand~\cite{marag2024,maoarag2024,lmars2024}. These paradigms underscore the importance of diagnostic signals that can inform decisions on whether to continue ranking, move on to refinement, or solicit human interaction.

Depth-wise analysis of relevance distributions can potentially provide the intervention signals required in an agentic workflow of information seeking. By observing how retrieval effectiveness evolves with increasing depth, an agent can determine if additional ranking or re-ranking is likely to surface further relevant information or whether one should instead route the query to an alternative agent (e.g., a reformulator ~\cite{jiang2018rin,hred_attention,HassanAwadallah,Muntean:2013,querysuggestion,DBLP:conf/cikm/PalumboDWLFFFSB23} or clarification agent~\cite{aliannejadi2019asking,zamani2020clarifying,mo2023convgqr}) or involve human-in-the-loop engagement~\cite{mo2023convgqr}.


\begin{figure*}[t]
\centering
\includegraphics[width=0.9\textwidth]{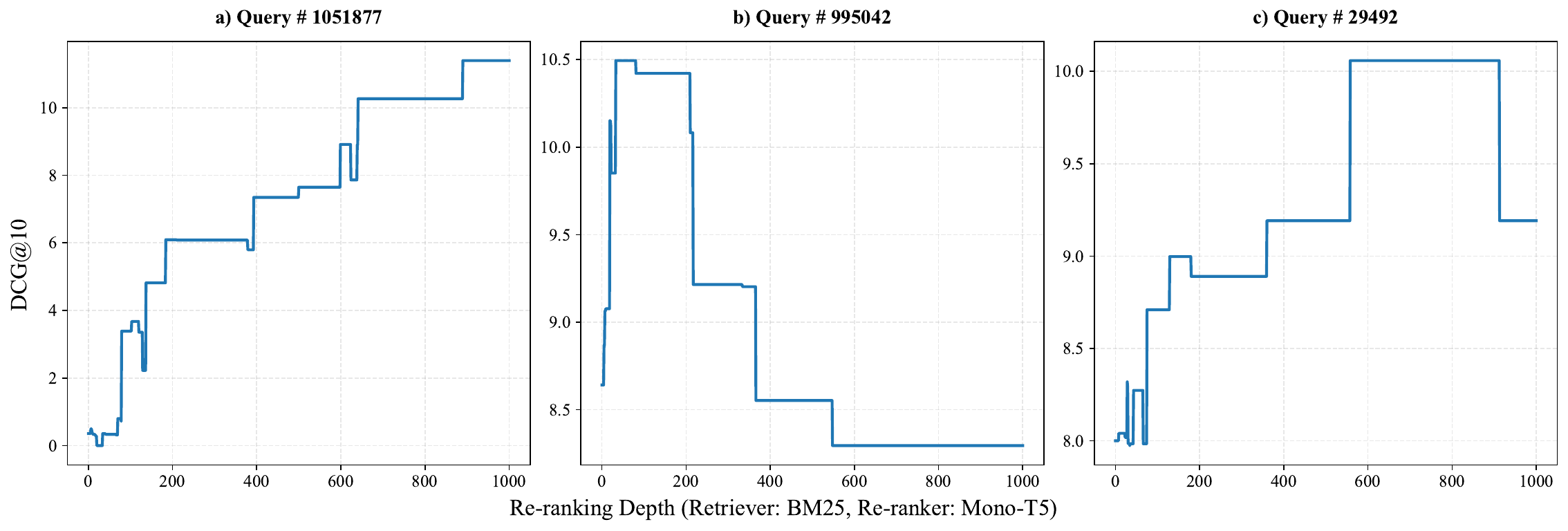}
\caption{a) For query ``\texttt{who sings the song 8675 jenny}'' (id: 1051877) we observe that as re-ranking depth increases, metrics (e.g., DCG@10) generally improve; b) For query ``\texttt{where is the clackamas county fair held?}'' (id: 995042) and query ``\texttt{attorney general massachusetts phone number}'' (id: 29492) increasing re-ranking depth does not consistently improve retrieval quality and may even lead to performance degradation.} ~\label{fig:per_query_performance} 
\end{figure*}



We argue that identifying such failure cases for individual queries requires explicit understanding of \emph{where relevance resides within ranked lists}. In particular, determining whether a query is limited by ranking quality or by query formulation depends on observing how relevant documents are distributed across retrieval depths (Figure~\ref{fig:fixed_depth}).
Importantly, this distribution is not only a property of the ranking model itself, but a latent characteristic of both the query and the documents within a collection that are relevant to it. 

Conducting this type of depth-wise analysis has been feasible only for a small number of queries, due to the high cost and limited coverage of human relevance judgments. Recent advances in large language models (LLMs) has now made it possible to apply these models as automated relevance assessors, producing query-specific relevance judgments at scale with quality comparable to human annotations~\cite{llmhumanrelevance,llmsurvey,llmforIRsurvey}. This motivates us to conduct a \textit{large-scale per-query depth-wise analysis of relevance distributions}, making it possible to \textit{identify retrieval failure cases} that were previously difficult to observe. Such analysis allows provision to directly observe where relevant documents appear, when retrievable relevance is exhausted, and whether additional retrieval or re-ranking effort yields any further gains.

Using dense LLM-based judgments, we directly observe depth--effectiveness trajectories for each query. As shown in Figure~\ref{fig:per_query_performance}, queries exhibit markedly different behaviors: some benefit monotonically from deeper re-ranking, others plateau early, and some experience clear degradation as depth increases. For each query, we define an optimal re-ranking depth as the shallowest depth that achieves maximal effectiveness under the ideal DCG envelope. We also analyze an LLM-assessed estimate of first-stage ranking quality (IR-DCG@10) as an indicator of whether deeper re-ranking is likely to be beneficial. Importantly, although both quantities (IR-DCG \& downstream effectiveness) derive from LLM assessments, they are computed over different objects---first-stage versus re-ranked lists.

\raggedbottom
\noindent{\bf Our contributions include:}

\uls
\li Our analysis reveals that queries exhibit distinct optimal re-ranking depths, and that re-ranking beyond this depth can degrade effectiveness below that of the initial ranking.

\li We evaluate a suite of standard query performance prediction (QPP) methods as proxies for selecting re-ranking depth and find that they are generally ineffective under this setting as compared to ideal-DCG computed with the automated judgments.

\li We show that an LLM-assessed estimate of first-stage ranking quality (IR-DCG@10) is more informative than standard QPP for calibrating depth, reducing average re-ranking depth by up to a factor of three while preserving overall effectiveness across multiple first- and second-stage configurations.

\ule


To facilitate further research, we release 53,936 LLM-based relevance judgments over MSMARCO v1 DEV and 5,285 judgments over TREC DL 2019, along with code and other artefacts\footnotemark[1].

\footnotetext[1]{\href{https://anonymous.4open.science/r/optimal_reranking_depth-717E}{\url{https://anonymous.4open.science/r/optimal_reranking_depth-717E}}}


\section{Related Work}
\label{sec:related}

\para{Ranked List Truncation (RLT)}
RLT methods~\cite{RLT,fixedRLT,variableRLT,RLTLLMrerank, pEBR} aim to determine a cut-off point $k$ for a ranked list based on retrieval signals, either globally or on a per-query basis. RLT methods are typically designed as reactive stopping mechanisms, optimizing a trade-off between effectiveness and cost under incomplete relevance judgments.

Our study is conceptually distinct. Rather than deciding when to stop re-ranking based on observed signals, we analyze re-ranking depth as a query-intrinsic property under near-complete relevance information. This allows us to ask a different question: whether the optimal re-ranking depth of a query can be predicted \textit{a priori} from properties of its first-stage retrieval under dense relevance judgments, and how much effectiveness and efficiency are theoretically recoverable under such predictions.

\para{Query Performance Prediction (QPP)}
QPP methods aim to estimate the quality retrieved results without requiring relevance judgments. Motivated by previous work~\cite{vdp} demonstrating QPP models as decent indicators of the pool depth for relevance assessments, we explore whether they can be effective predictors of the optimal re-ranking depth. Because QPP measures try to predict the difficulty of a query, a difficult query may require a deeper re-ranking depth. To explore this potential, we examine 11 representative predictors.
NQC~\cite{nqc} and its generalised version, SCNQC~\cite{ScaledNQC}, use the variance of the retrieval scores as an approximation of retrieval quality. 
QV-NQC~\cite{QVNQC} aggregates NQC estimates on a number of variants of the original input query. Both $\sigma_\text{max}$~\cite{perez2010standard}
and
$n(\sigma_{x\%})$~\cite{sigma} involve computing the retrieval score variance, the former using the maximum of score variance over an iteratively increasing set of top-retrieved documents, while the latter computes the variance on a fraction $x$ of the highest-ranked document's score.
SMV~\cite{SMV} uses the average and variance of the document scores as a predictor. WIG~\cite{WIG} measures the aggregated value of the information gain of each top-retrieved document with respect to the collection. DenseQPP~\cite{denseQPP} measures the diameter of the embedded representations of the top-ranked documents -- a smaller value indicating lower diversity and likely a better retrieval performance. UEF~\cite{UEF} computes a weighted average of QPP estimates over sublists sampled from the top-retrieved document list, the weights of these lists computed by the stability of relevance feedback. RSD~\cite{RSD} utilises bootstrap sampling to mitigate the impact of extreme values on retrieval scores. OddsRatio~\cite{OddsRatio} examines the consistency of pairwise comparison likelihoods.

\section{Methodology}

We formulate re-ranking depth selection as a \emph{predictive, query-specific} problem. Given a query $q$, a first-stage retriever produces a ranked list of candidate documents, which is subsequently re-ranked to depth $k$ by a more expensive model. Instead of fixing $k$ globally across all queries, we seek to predict a query-dependent depth $\hat{d}(q)$ \emph{a priori} from properties of the first-stage ranking, such that retrieval effectiveness is preserved while average re-ranking cost is minimized.

\para{Optimal Re-Ranking Depth}
\label{sec:dcg}

Let $\mu(k)$ denote the DCG@10 obtained after re-ranking the top $k$ documents:
\[
\mu(k) \;=\; \sum_{i=1}^{10} \frac{rel_i^{(k)}}{\log_2(i+1)},
\]
where $rel_i^{(k)} \in \{0,1,2,3\}$ denotes the relevance label at rank $i$ under re-ranking depth $k$. For each query, we define the maximum attainable effectiveness
\[
\mu^\star(q) \;=\; \max_k \mu(k),
\]
and the \emph{optimal re-ranking depth}
\[
d^\star(q) \;=\; \min \{ k : \mu(k) = \mu^\star(q) \}.
\]
This definition captures the shallowest depth at which a query attains its best possible effectiveness, thereby isolating unnecessary re-ranking beyond that point. 
Figure~\ref{fig:fixed_depth} plots $d^\star(q)$ averaged over all queries, as the ``optimal'' point. 
Our objective is to predict a depth $\hat{d}(q)$ such that $\mu(\hat{d}(q))$ closely approximates $\mu^\star(q)$ while minimizing the expected re-ranking cost $\mathbb{E}_q[\hat{d}(q)]$.

\para{Depth Predictors}

We consider depth predictors that assign each query a scalar score $\phi(q)$ intended to reflect the quality or difficulty of its first-stage retrieval. Rather than training models to regress re-ranking depth directly, which would tightly couple predictors to a specific depth distribution, we adopt a calibration-based approach that maps predictor values to depths in a non-parametric manner. 

We first experiment with established QPP methods as predictors. We also propose our own predictor. 
We compute DCG@10 over the initial retrieved list using LLM-based relevance judgments, yielding \textbf{IR-DCG@10}. 

\para{Percentile-Based Depth Calibration}

We map predictor values to re-ranking depths using percentile-based calibration. Under $K$-fold cross-validation, training queries are partitioned into equal-frequency bins according to $\phi(q)$. Within each bin, we collect the optimal depths $\{d^\star(q)\}$ of the training queries.

For a test query, we identify the corresponding bin and predict
\[
\hat{d}(q; p) \;=\; \text{$p$th percentile of } \{d^\star(q)\},
\]
where $p \in [0,1]$ explicitly controls the efficiency--effectiveness trade-off. Smaller values of $p$ bias predictions toward shallower re-ranking, prioritizing efficiency, while larger values favor deeper re-ranking to preserve effectiveness. This calibration mechanism decouples depth prediction from any single operating point.

\para{Operating Point Selection}

We sweep $p \in [0,1]$ and select the smallest value $p_{\min}$ satisfying
\[
\mathbb{E}_q \!\left[ \frac{\mu(\hat{d}(q; p))}{\mu(1000)} \right] \;\ge\; 1.
\]
This criterion identifies the shallowest operating point at which predicted depths achieve, on average, effectiveness equal to or exceeding that of fully re-ranking all queries to depth 1000. If parity is not reached, we report the value of $p$ yielding the highest observed effectiveness ratio.

For ease of reporting and deployment, we convert the percentile $p_{\min}$ to a threshold in the predictor score space. Specifically, once $p_{\min}$ is fixed, it induces a corresponding predictor threshold
\[
\texttt{thr}_{\min} \;=\; \text{$p_{\min}$-quantile of } \{ \phi(q) \}_{q \in \mathcal{Q}_{\mathrm{train}}}.
\]
At inference time, $\texttt{thr}_{\min}$ serves as an equivalent representation of the chosen operating point: it is the predictor-space cutoff that implements the depth predictions associated with $p_{\min}$.

\section{Evaluation}

\para{Datasets} We retrieve from the MSMARCO v1 passage corpus~\cite{msmarco}, a collection of 8.8 million segmented web documents mined from Bing query logs. To allow more granular exploration of ranking depth than is possible under existing human-annotated collections, we sample 1000 queries from the MSMARCO passage v1 dev set and annotate the top-10 rankings of multiple retrievers as a shallow judgment pool. In particular, we applied the UMBRELA framework for annotation~\cite{umbrela} using GPT-4o-mini as the underlying LLM. We also replicate the experiment on TREC DL 2019~\cite{dl2019}.

\begin{figure}[t]
\includegraphics[width=\columnwidth]{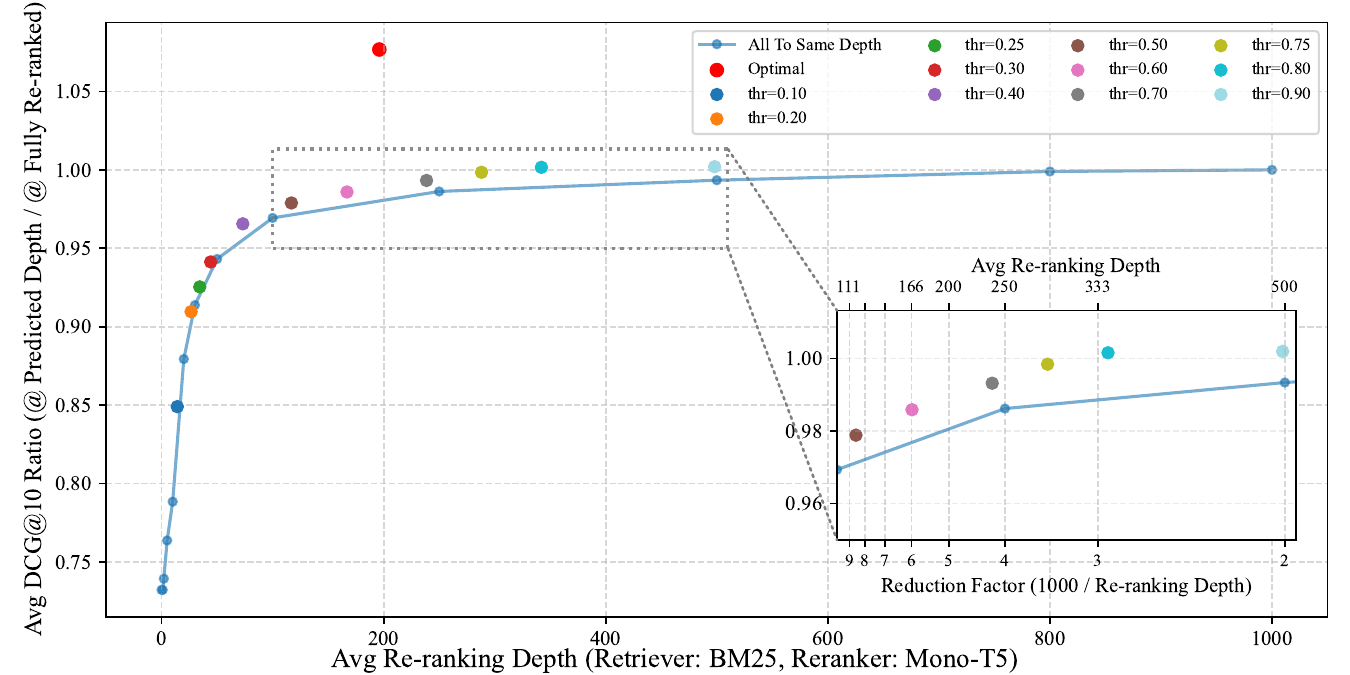}
\caption{Performance when using the IR-DCG@10 value as the predictor. The bottom-right part of the figure shows a close-up view of performance for $thr \ge 0.5$. At $thr = 0.8$, the average re-ranking depth is reduced by a factor of 3 while improving effectiveness.} \label{fig:initial_dcg_as_predictor_performance}
\end{figure}

\para{Experiment Setup} We evaluate retrieval effectiveness using DCG@10 as defined in Section \ref{sec:dcg}, following common practice on MSMARCO-style test collections~\cite{craswell:2019, craswell:2020}.
As retrievers, we apply BM25~\cite{bm25} ($k_1~=~1.2$, $b~=~0.75$) and the learned sparse model SPLADE~\cite{splade}. As re-rankers, we employ cross-encoders Mono-T5~\cite{monot5} and MonoELECTRA~\cite{MonoElectra}.
As depth predictors, we compute IR-DCG@10 and the 11 QPP approaches described in Section~\ref{sec:related}.

\para{Results}


Our first experiment uses BM25 as the retriever and Mono-T5 as the re-ranker. Table~\ref{tab:qpp_measures} summarizes the performance of each predictor. As shown in Table~\ref{tab:qpp_measures}, most QPP methods are ineffective in predicting the optimal re-ranking depth. From Table~\ref{tab:qpp_measures} The best-performing QPP baseline is RSD, which at $thr = 0.89$ reduces depth by a factor of 2 while maintaining full re-ranking effectiveness.

Our primary predictor is the IR-DCG@10 value derived from automated relevance assessments. From Figure~\ref{fig:initial_dcg_as_predictor_performance}, we can see that using IR-DCG@10 as the predictor and setting $thr=0.80$ can reduce the average re-ranking depth by a factor of 3, while maintaining overall retrieval effectiveness. We replicated the experiment using the TREC DL 2019 benchmark, observing a similar performance trajectory. Although both IR-DCG@10 and the target metric use LLM-derived judgments, they are computed over disjoint objects (first-stage vs \ re-ranked lists), all QPP baselines share the same labels yet perform far worse, and are consistent with TREC DL 2019\footnote{For brevity, these results can be found in our repository.}. 

Using SPLADE, observe in Figure~\ref{fig:fixed_depth}a, that there is little benefit in deep re-ranking SPLADE results with Mono-T5, since the optimal average re-ranking depth is around 20. Our findings align with previous work~\cite{RLTLLMrerank}, suggesting that if the retriever is already quite effective, a fixed re-ranking depth of around 20 is usually adequate.

\begin{table}[t]
\centering
\begin{tabular}{@{}lrrr@{}}
\toprule
\textbf{Predictor} & $\text{thr}_{min}$  &
$\mathbb{E}_q[\frac{\mu(\hat{d}(q))}{\mu(1000)}]$ &
\textbf{$\mathbb{E}_q[\hat{d}(q)]$} ~\\
\midrule
Dense-QPP~\cite{denseQPP} & 1.00 & 0.99 & 992.38~\\
WIG~\cite{WIG} & 1.00 & 0.99 & 961.30~\\
Odds Ratio~\cite{OddsRatio} & 1.00 & 1.00 & 923.64~\\
$\sigma_{\text{max}}$~\cite{perez2010standard} & 0.99 & 0.99 & 894.31~\\
NQC~\cite{nqc} & 0.98 & 1.00 & 846.44~\\
QV-NQC~\cite{Oleg_2019,QVNQC} & 0.98 & 1.00 & 846.44~\\
SCNQC~\cite{ScaledNQC} & 0.98 & 1.00 & 846.44~\\
$n(\sigma_{x\%})$~\cite{sigma} & 0.98 & 0.99 & 834.60~\\
SMV~\cite{SMV} & 0.97 & 1.00 & 764.22~\\
UEF~\cite{UEF} & 0.96 & 1.00 & 736.39~\\
RSD~\cite{RSD} & 0.89 & 1.00 & 511.29~\\
\textbf{IR-DCG@10} (ours) &~\textbf{0.79} &~\textbf{1.00} &~\textbf{331.26}~\\
\bottomrule
\\
\end{tabular}
\caption{The smallest threshold ($\text{thr}_{min}$) at which each predictor outperforms the retrieval quality obtained by re-ranking all queries to depth 1000. We also report the average ratio of each query’s DCG@10 at depth $d$ to its DCG@10 when fully re-ranked (depth 1000).
If the ratio never reaches 1, the threshold with the highest observed ratio is reported instead. The column $\mathbb{E}_q[\hat{d}(q)]$ reports the average predicted depth.
}
\label{tab:qpp_measures}
\vspace*{-\baselineskip}
\end{table}

We also experiment with MonoELECTRA on top of BM25 and SPLADE. Results and plots are also included in our repository. MonoELECTRA exhibits superior compatibility with SPLADE as performance is mostly asymptotic. Consistent with previous findings, under oracle selection of per-query optimal depths, we observe that the average re-ranking depth can be reduced by a factor of 3 for BM25 and by a factor of 50 for SPLADE (Figure~\ref{fig:fixed_depth} b). Our proposed predictor, IR-DCG@10, achieves a reduction factor of approximately 15 while matching the effectiveness of fully re-ranking. For BM25, our method generally outperforms fixed-depth re-ranking at the same average re-ranking depth.
\section{Conclusion}

Dense LLM-based relevance judgments enable us to study re-ranking depth at a scale infeasible with manual annotation. Using these labels, we show that re-ranking depth is query-specific: deeper re-ranking does not monotonically improve effectiveness, and standard QPP signals do not reliably predict optimal depth. Our analysis in an oracle setting reveals a remaining gap, indicating that predicting optimal depth is both an efficiency and an effectiveness problem. More broadly, this work highlights automated relevance assessment as an enabling methodology for studying adaptive retrieval behaviour at scale. Additionally applying our proposed measure (IR-DCG@10) derived from an LLM judge, we show that re-ranking depth can be predicted a priori with a calibrated predictor for a given first-stage. This predictor reduces average re-ranking depth by a factor of 3 while preserving effectiveness. We encourage future work to leverage dense automated judgments to develop stronger predictors, moving toward more efficient and query-adaptive retrieval pipelines.



\bibliographystyle{ACM-Reference-Format}
\bibliography{bib,proactive}

\end{document}